\documentclass[11pt,a4paper]{article}

\usepackage[utf8]{inputenc}
\usepackage[T1]{fontenc}
\usepackage{lmodern}
\usepackage[margin=2.6cm]{geometry}
\usepackage{amsmath,amssymb,amsthm}
\usepackage{graphicx}
\usepackage{booktabs}
\usepackage{array}
\usepackage{enumitem}
\usepackage[square,authoryear]{natbib}
\usepackage{tikz}
\usetikzlibrary{decorations.pathreplacing,arrows.meta,calc}
\usepackage[hidelinks]{hyperref}
\usepackage{url}

\graphicspath{{figures/}}

\newtheorem{proposition}{Proposition}
\newtheorem{remark}{Remark}

\newcommand{\lt}[1]{#1^{l}}
\newcommand{\rt}[1]{#1^{r}}
\newcommand{\FVect}{\mathbf{FVect}}
\newcommand{\Preg}{\mathbf{P}}
\newcommand{\id}{1}
\newcommand{\ip}[2]{\langle #1 \,|\, #2\rangle}

\title{A Categorical Compositional Distributional Modelling\\ for the Language of Life\thanks{Version~4 (18~November~2026). This version corrects the composition of protein meanings (the linker tensor, the meaning space of proteins, proteins with more than two domains), aligns the categorical set-up with \citet{Coecke2010}, corrects several biological and bibliographic details, and adds a section on developments since 2019. Versions~1--3 appeared in February, April and August~2019, co-authored with Quanlong Wang; from this version the paper has a single author (see Acknowledgements).}}

\author{Yanying Wu\\[1ex]
\small Centre for Neural Circuits and Behaviour, University of Oxford, Oxford, United Kingdom\\
\small \texttt{yanying.wu@cncb.ox.ac.uk}}

\date{18 November 2026}

\begin{document}
\maketitle

\begin{abstract}
The Categorical Compositional Distributional (DisCoCat) model is a mathematical framework for computing the meaning of a sentence from the meanings of its words and the grammatical structure that combines them. Because biological sequences can be viewed as a ``language of life'', it is tempting to ask whether the DisCoCat model can also give insight into biological macromolecules. Here we take a first step in this direction for proteins, whose linguistic features are the most prominent among the macromolecules. We treat a protein as a sentence and its domains as words, take the biological function of a domain or protein as its meaning, and regard the arrangement of domains as the syntax. We give pregroup grammar types to protein domains and to the linker regions between them, show how type reductions are lifted to multilinear maps between vector spaces, and use these maps to compute a vector representation of a protein's function from vector representations of its domains. In particular, linkers act as relational ``words'' whose order-three tensors combine the two flanking domains into a vector in a single protein meaning space, so that the functions of proteins with different domain architectures can be compared by an inner product. We extend the construction to proteins with any number of domains, illustrate it with a numerical toy example and five \textit{Drosophila melanogaster} proteins, discuss how the linker tensors could be learned and how the model could be tested, and relate it to protein language models and other developments since 2019.
\end{abstract}

\section{Introduction}

The metaphor of the genome as a ``book of life'' arose naturally once it was found that the genetic information of all living organisms is encoded in linear strings of DNA (deoxyribonucleic acid) \citep{Searls2002}. In this view, the language of life is written in DNA sequences, with the bases A, C, G and T as its alphabet and genes playing the part of sentences. The human genome, for example, contains about 3~billion base pairs and about 20{,}000 protein-coding genes, so the ``book of human'' is about 3~billion letters long and contains over 20{,}000 ``sentences''. Sequencing has made almost every letter of this book readable, and decades of genetic and molecular research have revealed the function of thousands of genes. Nevertheless, the function of many genes is still unknown, and the meaning of the book remains, to a great extent, obscure.

Treating DNA sequences as a language is not only interesting but also useful, because it allows linguistic tools and theories to be applied to the genome.

When the double-helix structure of DNA was discovered in the 1950s, the linguist Noam Chomsky was developing his theory of the syntactic structure of natural languages \citep{Chomsky1957}. Building on Chomsky's theory, David Searls and others carried out extensive grammatical analyses of macromolecules: DNA, RNA (ribonucleic acid) and proteins \citep{Doerfler1982,Brendel1984,Searls1992,Searls1993,Searls1997,Searls2002,Chiang2006,Dyrka2009,Sciacca2011}. These studies gave many insights into the structural composition of macromolecules from a linguistic perspective.

Also in the 1950s, the mathematician Joachim Lambek introduced a syntactic calculus \citep{Lambek1958} to formalize categorial grammar, which differs fundamentally from Chomsky's generative model. The basic ideas of categorial grammar go back to Kazimierz Ajdukiewicz \citep{Ajdukiewicz1935} and Yehoshua Bar-Hillel \citep{BarHillel1953}. Categorial grammar is built on the principle of compositionality: syntactic constituents combine as functions and arguments. Lambek's calculus formalized the type constructors for such functions and the rules for combining them \citep{Lambek1958}, and Lambek developed categorial grammar further in later work \citep{Lambek1968,Lambek1969,Lambek1972,Lambek1988,Lambek1999}. Building on Lambek's work, Bob Coecke and colleagues constructed a mathematical model of natural-language meaning called the Categorical Compositional Distributional (DisCoCat) model \citep{Coecke2010}. The DisCoCat model computes the meaning of a sentence by combining the meanings of its words according to its grammatical structure, and it has received experimental support on real corpus data \citep{Grefenstette2011}. It is therefore tempting to bring the DisCoCat model to macromolecular linguistics, so that we may understand not only the structure but also, more importantly, the meaning, that is the function, of these basic biological entities.

In this work we first review the linguistic analyses of macromolecules built on Chomsky's framework (Section~\ref{sec:chomsky}). We then introduce Lambek's pregroup grammar (Section~\ref{sec:pregroup}) and the DisCoCat model (Section~\ref{sec:discocat}), and develop a DisCoCat model for proteins (Section~\ref{sec:protein}). We focus on proteins rather than DNA or RNA because their modular, domain-based organisation makes them the easiest starting point. Section~\ref{sec:recent} relates the model to developments since the first version of this work (2019), and Section~\ref{sec:discussion} discusses its limitations and how it could be tested.

\section{Macromolecular linguistics based on Chomsky's framework}\label{sec:chomsky}

Perhaps because of its dominant position in linguistics, Chomsky's generative grammar, rather than Lambek's type grammar, has been the framework of choice for researchers studying the language of life. In this section we briefly review what has been achieved in macromolecular linguistics within Chomsky's framework.

Although DNA carries the hereditary information, it is not the only macromolecule with a linear sequence. RNA molecules and proteins also have intrinsic linear sequences, and they play indispensable roles in all living organisms.

\begin{figure}[htbp]
\centering
\includegraphics[width=0.6\textwidth]{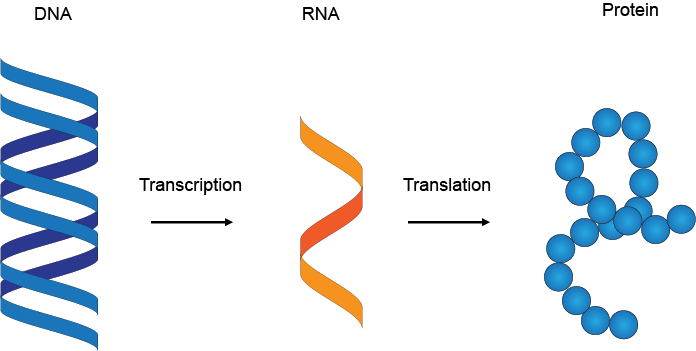}
\caption{The central dogma of molecular biology: DNA is transcribed into RNA, which is translated into protein.}
\label{fig:dogma}
\end{figure}

According to the central dogma of molecular biology (Figure~\ref{fig:dogma}), the information in DNA is first transcribed into RNA and then translated into protein. It is the RNA molecules and the proteins that carry out the biological functions of genes.

The linguistic view of biological sequences has focused on grammar-like structures in genes, RNA molecules and proteins. We briefly review the relevant molecular structures and then describe their analyses with formal grammars.

\subsection{The biological structures of genes, RNA molecules and proteins}

\begin{figure}[htbp]
\centering
\includegraphics[width=0.5\textwidth]{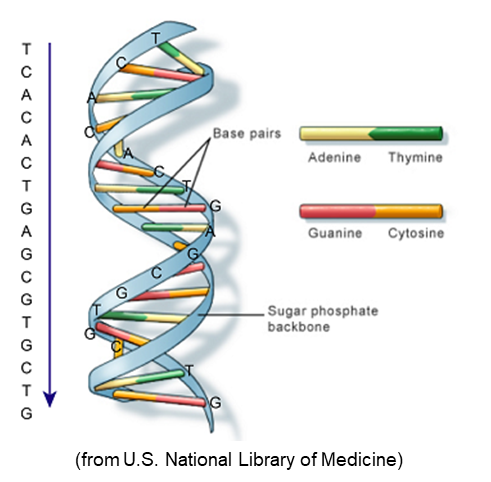}
\caption{The structure and sequence of DNA (image from the U.S.\ National Library of Medicine).}
\label{fig:dna}
\end{figure}

DNA is the hereditary material that carries the genetic information of humans and almost all other organisms. The information in DNA is stored as a code made up of four chemical bases: adenine (A), guanine (G), cytosine (C) and thymine (T). Human DNA consists of about 3~billion bases, and more than 99~percent of these bases are the same in all people \citep{MedlinePlusDNA}. The bases pair with each other, A with T and C with G, to form base pairs. Each base is attached to a sugar and a phosphate to form a nucleotide, and the nucleotides are arranged in two long strands that wind around each other to form a double helix. The sequence of a DNA molecule is written as the sequence of bases along one strand (Figure~\ref{fig:dna}).

\begin{figure}[htbp]
\centering
\includegraphics[width=0.5\textwidth]{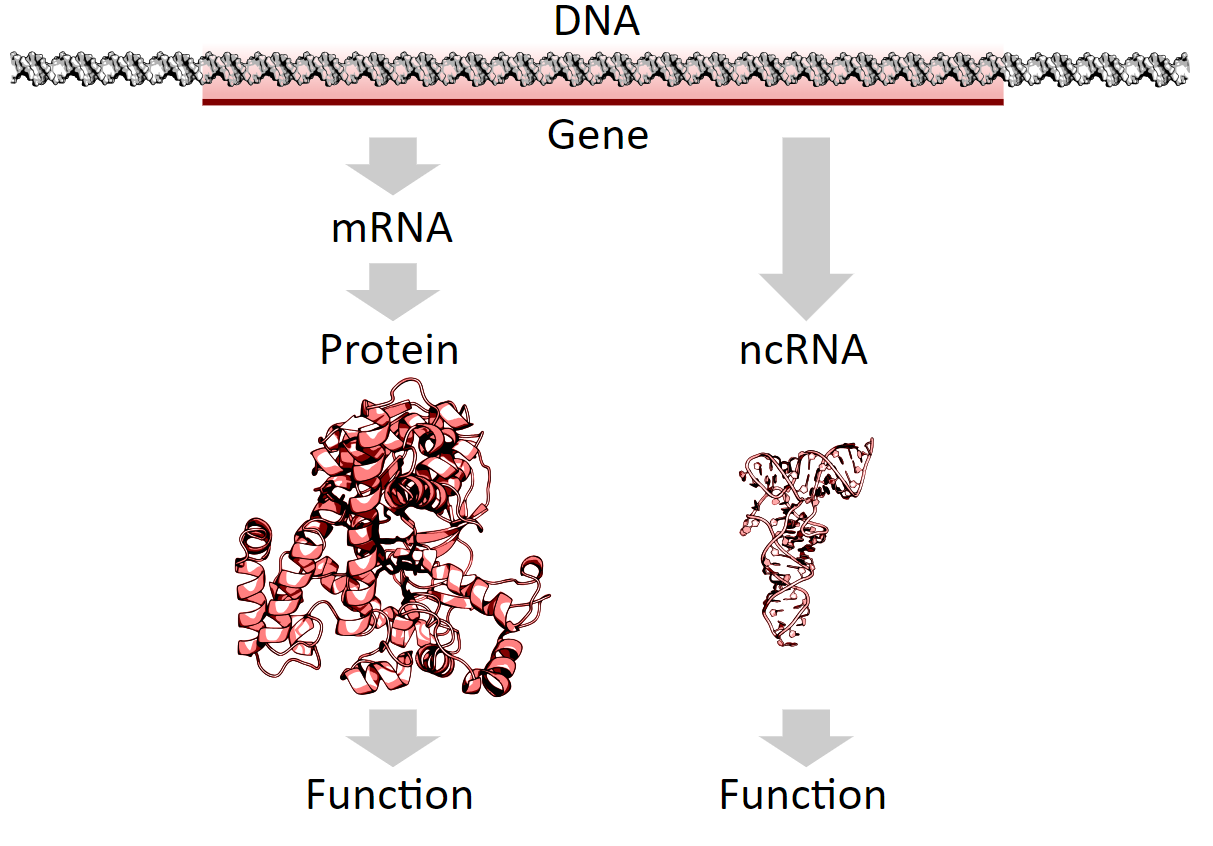}
\caption{Functional expression of a gene as a protein (via mRNA) or as a non-coding RNA (ncRNA) (image by Thomas Shafee, CC~BY, via Wikimedia Commons).}
\label{fig:expression}
\end{figure}

A gene is a segment of DNA that can be expressed as a functional RNA or protein (Figure~\ref{fig:expression}). A protein-coding gene is transcribed into a messenger RNA (mRNA), which is then translated into a protein with a specific biological function, as stated by the central dogma. Other genes are transcribed into non-coding RNAs (ncRNAs), such as transfer and ribosomal RNAs, which are not translated but have biological functions of their own. Besides their primary structure, that is the linear sequence, both proteins and ncRNAs have secondary structure: local folding patterns stabilized by hydrogen bonds. The most common secondary structures in proteins are the $\alpha$-helix and the $\beta$-sheet; in RNA they are stem-loops (hairpins), which can combine into higher-order motifs such as pseudoknots.

\subsection{A brief review of macromolecular linguistics}

\begin{figure}[htbp]
\centering
\includegraphics[width=0.75\textwidth]{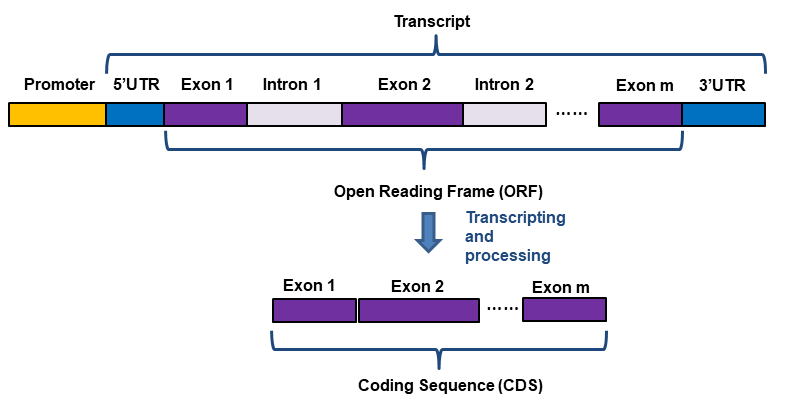}
\caption{A typical eukaryotic gene. The transcript contains the 5$'$~UTR, exons separated by introns, and the 3$'$~UTR. Splicing removes the introns; the coding sequence (CDS) is the translated part of the spliced mRNA. UTR, untranslated region; ORF, open reading frame.}
\label{fig:gene}
\end{figure}

We begin with the grammar of genes. A typical eukaryotic protein-coding gene has the structure shown in Figure~\ref{fig:gene}. It consists of a promoter followed by a transcribed region. The primary transcript contains a 5$'$ untranslated region (5$'$~UTR), a series of exons separated by introns, and a 3$'$~UTR. During RNA processing the introns are removed by splicing. The UTRs remain in the mature mRNA but are not translated; the part of the mRNA that is translated into protein is the coding sequence (CDS), which forms an open reading frame (ORF) running from a start codon to a stop codon. In Figure~\ref{fig:gene}, the bracket labelled ORF marks the genomic span from the first to the last coding exon; strictly speaking, the ORF is defined on the spliced mRNA.

Like a sentence, such a gene has a modular structure, which can be captured by a set of grammar rules \citep{Searls2002,Searls2013}:
\begin{align*}
\textit{Gene} &\to \textit{Promoter}\ \textit{Transcript}\\
\textit{Transcript} &\to \textit{5$'$UTR}\ \textit{ORF}\ \textit{3$'$UTR}\\
\textit{ORF} &\to \textit{Exon}\ \textit{Intron}\ \textit{ORF} \mid \textit{Exon}\\
\textit{Exon} &\to \textit{CodingSeq}\\
\textit{Intron} &\to \textit{Donor}\ \textit{SkippedSeq}\ \textit{Acceptor}\\
\textit{Donor} &\to \texttt{gt}\\
\textit{Acceptor} &\to \texttt{ag}\\
\textit{CodingSeq} &\to \texttt{atgaacaca}\ldots\texttt{taa}\\
\textit{SkippedSeq} &\to \texttt{cccaagc}\ldots\texttt{gca}
\end{align*}
Here \textit{CodingSeq} and \textit{SkippedSeq} play the role of words in a sentence; the DNA sequences given for them are arbitrary placeholders with no real meaning. (In this simplified grammar, the ORF non-terminal stands for the exon--intron region of the gene, as in Figure~\ref{fig:gene}.) Much more elaborate gene grammars can be found in \citet{Dong1994} and \citet{Searls1993}. Such grammars can be used to recognize and predict the structures of genes that have not yet been characterized \citep{Dong1994}.

Next we turn to RNA secondary structure. Unlike DNA, whose two complementary strands pair along their whole length, an RNA molecule usually consists of a single strand. Nevertheless, RNA molecules often fold back on themselves into secondary structures such as hairpins (stem-loops), internal loops, bulges and pseudoknots.

Figure~\ref{fig:hairpin} shows a hairpin formed by the RNA sequence \texttt{GAUC\,GAAA\,GAUC}. Its stem is formed by the inverted repeat \texttt{GAUC}\,\ldots\,\texttt{GAUC} (the second copy is the reverse complement of the first), and \texttt{GAAA} forms the loop. Hairpins of this kind are generated by the context-free grammar
\begin{align*}
\textit{Stem} &\to \texttt{a}\ \textit{Stem}\ \texttt{u} \mid \texttt{u}\ \textit{Stem}\ \texttt{a} \mid \texttt{c}\ \textit{Stem}\ \texttt{g} \mid \texttt{g}\ \textit{Stem}\ \texttt{c} \mid \textit{Loop}\\
\textit{Loop} &\to N\,N\,N\ \textit{Tail}, \qquad \textit{Tail} \to N\ \textit{Tail} \mid \varepsilon, \qquad N \to \texttt{a} \mid \texttt{c} \mid \texttt{g} \mid \texttt{u}
\end{align*}
\citep[adapted from][]{Searls2013}, in which the first rule produces the nested base pairs of the stem and the \textit{Loop} rules produce an unpaired loop of at least three nucleotides, the minimum that is sterically possible. (A rule $\textit{Stem}\to\varepsilon$ alone, without a loop, would generate perfectly paired palindromes, which cannot fold into a hairpin. For simplicity, the grammar omits G--U wobble pairs.) More complex RNA secondary structures and their grammars can be found in \citet{Chiang2006} and \citet{Searls1999}.

\begin{figure}[htbp]
\centering
\begin{tikzpicture}[
  nt/.style={circle,draw,thick,minimum size=7mm,inner sep=0pt,font=\small\bfseries},
  bp/.style={line width=2.5pt,cyan!60!blue},
  bb/.style={thick,blue!60!black}]
\def\dx{1.6}
\node[nt,fill=orange!40] (n1) at (0,3.0) {G};
\node[nt,fill=red!25]    (n2) at (0,2.2) {A};
\node[nt,fill=violet!25] (n3) at (0,1.4) {U};
\node[nt,fill=green!25]  (n4) at (0,0.6) {C};
\node[nt,fill=orange!40] (n5) at (-0.2,-0.35) {G};
\node[nt,fill=red!25]    (n6) at (0.35,-1.05) {A};
\node[nt,fill=red!25]    (n7) at (1.25,-1.05) {A};
\node[nt,fill=red!25]    (n8) at (1.8,-0.35) {A};
\node[nt,fill=orange!40] (n9)  at (\dx,0.6) {G};
\node[nt,fill=red!25]    (n10) at (\dx,1.4) {A};
\node[nt,fill=violet!25] (n11) at (\dx,2.2) {U};
\node[nt,fill=green!25]  (n12) at (\dx,3.0) {C};
\foreach \a/\b in {1/2,2/3,3/4,4/5,5/6,6/7,7/8,8/9,9/10,10/11,11/12} \draw[bb] (n\a) -- (n\b);
\foreach \a/\b in {1/12,2/11,3/10,4/9} \draw[bp] (n\a) -- (n\b);
\draw[bb] (n1.north) -- ++(0,0.4) node[above,font=\small] {5$'$};
\draw[bb,-{Stealth}] (n12.north) -- ++(0,0.5) node[above,font=\small] {3$'$};
\node[font=\small,anchor=west] at (2.4,1.8) {stem (4 bp)};
\node[font=\small,anchor=west] at (2.4,-0.7) {loop (\texttt{GAAA})};
\end{tikzpicture}
\caption{An RNA hairpin formed by \texttt{GAUC\,GAAA\,GAUC}: a four-base-pair stem closed by a four-nucleotide loop. (Redrawn for this version; the hairpin in versions~1--3 had an empty loop, which cannot form.)}
\label{fig:hairpin}
\end{figure}

Finally, we turn to protein grammars. Proteins have a larger ``alphabet'' of 20 amino acids, and their structures are much more diverse and complicated than those of RNA molecules, so there is as yet no simple grammar for whole proteins. Nevertheless, the linguistic properties of protein structure have been studied extensively, and elegant models exist at least for local structures. These include basic secondary-structure elements such as $\alpha$-helices and $\beta$-sheets, as well as many important ligand-binding sites.

\begin{figure}[htbp]
\centering
\includegraphics[width=0.3\textwidth]{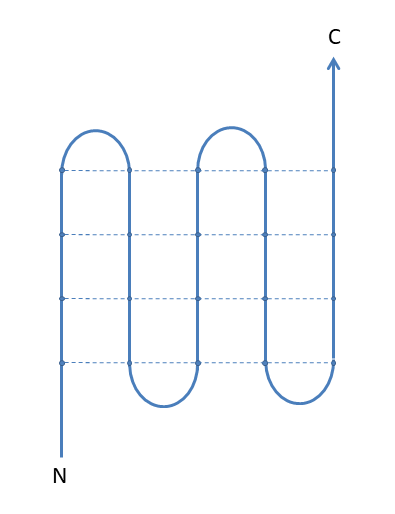}
\caption{A typical protein secondary structure: a $\beta$-sheet of five strands running from the N terminus to the C terminus.}
\label{fig:sheet}
\end{figure}

\begin{figure}[htbp]
\centering
\includegraphics[width=0.55\textwidth]{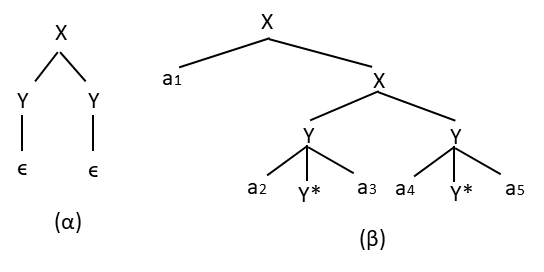}
\caption{Ranked node-rewriting grammar (RNRG) for a $\beta$-sheet of five strands: initial tree ($\alpha$) and auxiliary tree ($\beta$).}
\label{fig:rnrg}
\end{figure}

For example, Figure~\ref{fig:sheet} shows a typical $\beta$-sheet, and the ranked node-rewriting grammar (RNRG) in Figure~\ref{fig:rnrg} generates this $\beta$-sheet \citep{Chiang2006}.

In Figure~\ref{fig:rnrg}, $X$, $Y$ and $Y^{*}$ are non-terminal nodes, while $a_i$ ($i=1,\ldots,5$) and $\epsilon$ are terminal nodes (an amino acid and the empty string, respectively). An RNRG is essentially a tree-adjoining grammar (TAG) whose elementary trees may have several foot nodes. The grammar in Figure~\ref{fig:rnrg} has two elementary trees, ($\alpha$) and ($\beta$). The small tree ($\alpha$) on the left is an initial tree, whose root $X$ is the start node; the tree ($\beta$) on the right is an auxiliary tree, which has special foot nodes marked with~$*$.

\begin{figure}[htbp]
\centering
\includegraphics[width=0.6\textwidth]{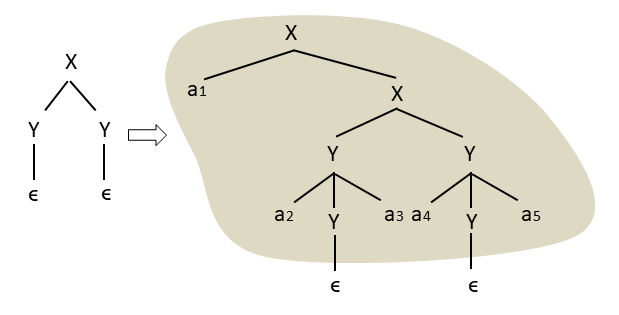}
\caption{Rewriting of node $X$ with the auxiliary tree ($\beta$) from Figure~\ref{fig:rnrg}.}
\label{fig:rewrite}
\end{figure}

In a regular or context-free grammar, such as those for DNA and RNA above, the basic operation is symbol rewriting: a symbol on the left-hand side of a rule is replaced by the right-hand side. In an RNRG, the basic operation is node rewriting. When a node $X$ is rewritten with a tree ($\beta$), the children of $X$ are identified with the foot nodes of ($\beta$), matched in order of linear precedence. Figure~\ref{fig:rewrite} shows a rewriting of node $X$ with tree ($\beta$) from Figure~\ref{fig:rnrg} \citep{Chiang2006}.

Many context-free grammars for protein structure can be converted into TAG and related formalisms \citep{Chiang2006}. Alternatively, stochastic context-free grammars can model protein ligand-binding sites efficiently \citep{Dyrka2009}.

\section{Lambek's pregroup grammar}\label{sec:pregroup}

In 1999, Lambek introduced pregroups as a new computational, algebraic approach to the syntactic structure of natural languages \citep{Lambek1999}.

A pregroup $(P,\to,\cdot,1,(-)^{l},(-)^{r})$ is a partially ordered monoid in which every element has a left and a right adjoint \citep{Lambek1999,Coecke2010}. Writing $\to$ for the partial order and juxtaposition for the monoid multiplication, a pregroup satisfies:
\begin{center}
\begin{tabular}{lll}
(1) & $(xy)z = x(yz)$ & (associativity)\\
(2) & $x1 = x = 1x$ & (identity)\\
(3) & $x \to x$ & (reflexivity)\\
(4) & if $x\to y$ and $y \to z$ then $x \to z$ & (transitivity)\\
(5) & if $x\to y$ and $y \to x$ then $x = y$ & (antisymmetry)\\
(6) & if $x\to y$ and $x' \to y'$ then $xx' \to yy'$ & (compatibility)\\
(7) & if $x \to y$ then $uxv \to uyv$ & (substitutivity)\\
(8) & $\lt{x}x \to 1$ and $x\rt{x} \to 1$ & (contractions)\\
(9) & $1 \to x\lt{x}$ and $1 \to \rt{x}x$ & (expansions)
\end{tabular}
\end{center}

The main idea of a pregroup grammar is to assign one or more types to each word of a language. The types are elements of a pregroup, so the grammatical status of a sentence, as a string of words, can be calculated from the types of its words \citep{Lambek1999,Lambek2008}.

To build such a pregroup, one starts from a partially ordered set $(B,\to)$ of basic types. The basic types stand for traditional grammatical categories such as noun or sentence, and for $x,y\in B$, $x \to y$ means that everything of type $x$ is also of type $y$. One then takes the free pregroup generated by $(B,\to)$: every basic type $p$ acquires a left adjoint $\lt{p}$ and a right adjoint $\rt{p}$ satisfying rules (8) and (9). Adjoints have the following properties \citep{Lambek1999}:
\begin{itemize}[nosep]
\item adjoints are unique;
\item adjoints are order-reversing: if $p \to q$ then $\rt{q}\to\rt{p}$ and $\lt{q}\to\lt{p}$;
\item the unit is self-adjoint: $\lt{1} = 1 = \rt{1}$;
\item adjoints reverse products: $(p q)^{r} = \rt{q}\rt{p}$ and $(p q)^{l} = \lt{q}\lt{p}$;
\item opposite adjoints cancel: $(\lt{p})^{r} = p = (\rt{p})^{l}$;
\item iterated adjoints of the same kind do not cancel: $p^{ll}\lt{p}\to 1\to \rt{p}p^{rr}$, $p^{lll}p^{ll} \to 1 \to p^{rr}p^{rrr}$, \ldots
\end{itemize}

Basic types and their iterated adjoints are called simple types, and a string $x_1\cdots x_n$ of simple types with $n\ge 0$ is a compound type; the empty string ($n=0$) is denoted by $1$. Strings are multiplied by concatenation:
\[
(x_1\cdots x_m)(y_1\cdots y_n) = x_1\cdots x_m y_1\cdots y_n .
\]
A string of words is a grammatical sentence if the concatenation of the types of its words reduces to the sentence type. By Lambek's switching lemma, such a reduction can always be carried out using contractions and the partial order alone, without expansions \citep{Lambek1999}.

This framework differs from Chomsky's in its algebraic, type-driven formulation, yet pregroup grammars are weakly equivalent to context-free grammars: they generate the same class of languages \citep{Buszkowski2001}.

When the first version of this paper appeared (2019), we were not aware of any description of macromolecular linguistics in terms of Lambek's pregroup grammar.

\section[The categorical compositional distributional (DisCoCat) model]{The categorical compositional distributional\\ (DisCoCat) model}\label{sec:discocat}

In 2010, \citet{Coecke2010} established a mathematical framework, the DisCoCat model, for computing the meaning of a sentence from the meanings of its words. The model uses category theory to unify distributional representations of word meanings, which are vectors, with the compositional grammatical types of a pregroup, and uses the pregroup reductions to transform the meanings of the words into the meaning of the whole sentence.

The key observation is that pregroups and vector spaces share the same high-level mathematical structure: both form compact closed monoidal categories \citep{Coecke2010,Grefenstette2011}.

\subsection{Monoidal and compact closed categories}

A strict monoidal category $\mathcal{C}$ consists of the following data \citep{Coecke2010}:
\begin{itemize}[nosep]
\item a collection of objects $A, B, C, \ldots$;
\item for each ordered pair of objects $(A,B)$, a set $\mathcal{C}(A,B)$ of morphisms, each written $f: A\to B$;
\item for $f: A\to B$ and $g: B\to C$, a composite $g\circ f: A\to C$, which is associative: $(h\circ g)\circ f = h\circ(g\circ f)$;
\item for each object $A$, an identity morphism $\id_A: A\to A$ with $f\circ \id_A = f = \id_B\circ f$ for every $f: A\to B$;
\item for each ordered pair of objects $(A,B)$, a composite object $A\otimes B$, the tensor product, which is associative, $(A\otimes B)\otimes C = A\otimes(B\otimes C)$, and has a unit object $I$ with $I\otimes A = A = A\otimes I$;
\item for each pair of morphisms $f: A\to C$ and $g: B\to D$, a parallel composite $f\otimes g: A\otimes B\to C\otimes D$, satisfying bifunctoriality:
\[
(g_1\otimes g_2)\circ(f_1\otimes f_2) = (g_1\circ f_1)\otimes(g_2\circ f_2).
\]
\end{itemize}
A morphism $\psi: I\to A$ is called an element, or state, of $A$; in the category of vector spaces these are exactly the vectors of $A$.

A monoidal category is compact closed if every object $A$ has a left adjoint $\lt{A}$ and a right adjoint $\rt{A}$, together with morphisms
\[
\eta^{l}: I\to A\otimes \lt{A},\qquad \epsilon^{l}: \lt{A}\otimes A\to I,\qquad
\eta^{r}: I\to \rt{A}\otimes A,\qquad \epsilon^{r}: A\otimes\rt{A}\to I,
\]
which satisfy
\begin{align*}
(\id_A\otimes\epsilon^{l})\circ(\eta^{l}\otimes\id_A) &= \id_A, & (\epsilon^{r}\otimes\id_A)\circ(\id_A\otimes\eta^{r}) &= \id_A,\\
(\epsilon^{l}\otimes\id_{\lt{A}})\circ(\id_{\lt{A}}\otimes\eta^{l}) &= \id_{\lt{A}}, & (\id_{\rt{A}}\otimes\epsilon^{r})\circ(\eta^{r}\otimes\id_{\rt{A}}) &= \id_{\rt{A}}.
\end{align*}
Because the tensor product need not be symmetric, $\lt{A}$ and $\rt{A}$ may differ; this non-symmetric case is the one relevant to grammar \citep{Coecke2010}.

\paragraph{Pregroups.} A pregroup is a compact closed category in which there is at most one morphism between any two objects (a posetal category): the objects are types, there is a morphism $x\to y$ exactly when $x\to y$ holds in the partial order, the tensor is concatenation, and the $\epsilon$ and $\eta$ morphisms are the contractions and expansions (8) and (9) \citep{Coecke2010}.

\paragraph{Vector spaces.} Let $\FVect$ be the category of finite-dimensional real vector spaces and linear maps, with the usual tensor product and $I=\mathbb{R}$. Here the tensor is symmetric and $\lt{V}=\rt{V}=V^{*}$; choosing an orthonormal basis $\{e_i\}_i$ identifies $V^{*}$ with $V$, and the compact closed structure is
\begin{equation}\label{eq:epsilon}
\eta^{l}=\eta^{r}: \mathbb{R}\to V\otimes V,\ 1\mapsto \sum_i e_i\otimes e_i,
\qquad
\epsilon^{l}=\epsilon^{r}: V\otimes V\to\mathbb{R},\ \sum_{ij} c_{ij}\, e_i\otimes e_j \mapsto \sum_i c_{ii}.
\end{equation}
On product vectors, $\epsilon$ is simply the inner product: $\epsilon(v\otimes w)=\ip{v}{w}$ \citep{Coecke2010}.

\subsection{From grammatical reductions to linear maps}\label{sec:lifting}

Once the common structure of vector spaces and pregroups is identified, grammatical reductions can be turned into linear maps. \citet{Coecke2010} do this in the product category $\FVect\times\Preg$, whose objects are pairs $(V,p)$ of a vector space $V$ (where the meanings of words of type $p$ live) and a pregroup type $p$, and whose morphisms are pairs $(f,\to): (V,p)\to(W,q)$ of a linear map $f: V\to W$ and an order relation $p\to q$. The compact closed structure lifts componentwise. For a string of words $w_1\cdots w_n$ with meaning spaces $(W_i,p_i)$ and a reduction $\alpha: p_1\cdots p_n\to x$, the linear map $f$ is obtained by replacing every type in $\alpha$ by the corresponding vector space, and the meaning of the string is
\begin{equation}\label{eq:meaning}
\overrightarrow{w_1\cdots w_n} := f(\vec w_1\otimes\cdots\otimes\vec w_n) \in X,
\end{equation}
where $X$ is the space of the reduced type $x$ \citep[Definition~3.2]{Coecke2010}.

Because a pregroup is posetal, it records only \emph{whether} a reduction exists, not \emph{which} reduction is used, whereas the map $f$ depends on the particular sequence of contractions. An equivalent and now common way to make this precise is to replace the pregroup by the free compact closed category generated by the basic types, whose morphisms are the reduction diagrams themselves \citep{Preller2007}, and to define a strong monoidal functor $F$ from it to $\FVect$ that sends each basic type to a vector space, each adjoint to the same space, each $\epsilon$ and $\eta$ to the maps in~\eqref{eq:epsilon}, and each order relation $x\to y$ to a chosen linear map $F(x)\to F(y)$ \citep{Coecke2013}. For the reductions used in this paper, both formulations give the same linear maps. (Versions 1--3 of this paper described the passage as a functor from the pregroup itself; the formulations above are the precise ones.)

Concretely, the meaning of a string of words $w_1 w_2\cdots w_n$ is computed in four steps \citep{Coecke2010}:
\begin{enumerate}[nosep]
\item assign a grammatical type $p_i$ to each word $w_i$;
\item the type determines the meaning space of the word, $W_i = F(p_i)$: basic types are sent to chosen spaces, and compound types to tensor products of these; take the meaning vector $\vec w_i \in W_i$ of each word, obtained from data;
\item concatenate the types into $p_1p_2\cdots p_n$, reduce this string with the rules of the pregroup to the target type, and record the reduction as a morphism $\alpha$;
\item lift $\alpha$ to a linear map $f=F(\alpha)$ and apply it to $\vec w_1\otimes\vec w_2\otimes\cdots\otimes\vec w_n$, as in~\eqref{eq:meaning}.
\end{enumerate}

\subsection{Comparing meanings}

An important property of the construction is that all grammatical sentences, whatever their structure, have meanings in the same space $S=F(s)$ of the sentence type $s$. Their meanings can therefore be compared with the normalised inner product (cosine similarity) \citep[Definition~5.1]{Coecke2010}:
\begin{equation}\label{eq:similarity}
\mathrm{sim}(\vec s,\vec s\,') = \frac{\ip{\vec s}{\vec s\,'}}{\lVert\vec s\rVert\,\lVert\vec s\,'\rVert}.
\end{equation}
This is the main payoff of the framework: it extends the distributional comparison of words to whole sentences. The structure also has a diagrammatic calculus, in which the $\epsilon$ maps are drawn as ``cups'' that connect the words of a sentence and show how information flows between them \citep{Coecke2010,Selinger2011}.

\section{A DisCoCat model for protein linguistics}\label{sec:protein}

The DisCoCat model combines distributional representations of word meanings with grammatical types to obtain the meaning of a sentence. We propose an analogous construction for proteins, in which a protein is a sentence and its domains and the linker regions between them are words. The function (meaning) of a protein is then obtained by combining the vector representations of its constituents according to their grammatical types.

\subsection{Why proteins are a good fit}

The modular nature of protein structure has long been recognized. To carry out a great diversity of biological functions, proteins fold into highly organized three-dimensional structures, and these structures are built from a limited repertoire of modules. According to the structural classification databases CATH (Class, Architecture, Topology, Homologous superfamily) \citep{Knudsen2010} and SCOP (Structural Classification of Proteins) \citep{LoConte2000}, there are only about 1{,}000--2{,}000 distinct folds (arrangements of secondary-structure elements), whereas the number of known protein structures and sequences is far larger. These data provide strong evidence for protein modularity \citep{Levy2017}.

Throughout evolution, the duplication and recombination of genes have been major sources of new proteins. The units that are recombined, and that themselves often remain largely unchanged, are protein domains: independently folding structural and functional units, each of which adopts one fold (many unrelated domain families share the same fold). Great efforts have been made to classify domains into families \citep{ElGebali2019,LoConte2000,Knudsen2010,Mitchell2019}, and how combinations of domains produce new functions has been studied intensively \citep{Apic2001,Moore2008,vonHeijne2018,Bashton2007,BornbergBauer2010,Levy2017}.

The analogy between the assembly of proteins from domains and the assembly of sentences from words has been pointed out explicitly \citep{Gimona2006,Bashton2007}. Beyond this general similarity, domain splitting deserves particular mention: in some cases during evolution, two functions previously carried out by a single-domain protein have been separated and taken over by two different domains of a new multi-domain protein. This separation of functions closely resembles the emergence of discreteness, a characteristic feature of the development of natural languages \citep{Senghas2004,Bashton2007}.

These similarities between proteins and natural languages motivate our attempt to apply the DisCoCat model to proteins. Doing so requires two ingredients: (1) a grammatical typing of proteins in a pregroup; and (2) vector representations of the meanings of domains. We prepare these in Sections~\ref{sec:typing} and~\ref{sec:vectors}.

\subsection{Grammatical typing of proteins}\label{sec:typing}

\begin{figure}[htbp]
\centering
\includegraphics[width=0.75\textwidth]{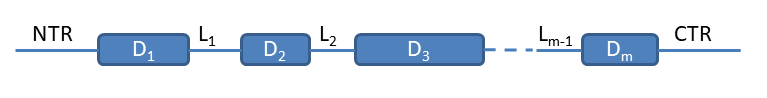}
\caption{A hypothetical protein: N-terminal region (NTR), domains $D_1,\ldots,D_m$, linker regions $L_1,\ldots,L_{m-1}$ and C-terminal region (CTR).}
\label{fig:hypo}
\end{figure}

We start from a hypothetical protein with the representative structure shown in Figure~\ref{fig:hypo}: an N-terminal region (NTR), domains $D_1,\ldots,D_m$ separated by linker regions $L_1,\ldots,L_{m-1}$, and a C-terminal region (CTR).

\paragraph{Basic types.} We type each domain by its broad functional class and introduce the following basic types:
\begin{center}
\begin{tabular}{ll@{\qquad}ll}
$b$ & binding domain & $c$ & catalytic domain\\
$b_1$ & ligand-binding domain & $c_1$ & oxidoreductase catalytic domain\\
$b_2$ & DNA-binding domain & $c_2$ & transferase catalytic domain\\
$b_3$ & RNA-binding domain & $c_3$ & hydrolase catalytic domain\\
$b_4$ & polypeptide-binding domain & $c_4$ & lyase catalytic domain\\
$i$ & protein--protein interaction domain & $c_5$ & isomerase catalytic domain\\
$r$ & regulatory domain & $c_6$ & ligase catalytic domain\\
$t$ & channel or transporter domain & & \\[0.5ex]
$p$ & protein & $p_1$, $p_2$ & single-domain, multi-domain protein
\end{tabular}
\end{center}
The catalytic subtypes $c_1,\ldots,c_6$ follow the six classic Enzyme Commission (EC) classes; the seventh class, translocases (EC~7), introduced in 2018 \citep{Tipton2018}, overlaps with our transporter type $t$. Versions~1--3 also listed basic types $n$ (N-terminal region), $\bar c$ (C-terminal region) and $l$ (linker), which were never used; in this version linkers receive compound types, and terminal regions are treated in Remark~\ref{rem:terminal}.

\paragraph{Order postulates.} Following Lambek's protocol \citep{Lambek2008}, we impose a partial order on the basic types; further relations follow by transitivity:
\begin{enumerate}[label=(P\arabic*),nosep]
\item $b_k\to b$ $(k=1,\ldots,4)$ and $c_j\to c$ $(j=1,\ldots,6)$: each subtype is an instance of its general type;
\item $b\to r$ and $i\to r$: binding and interaction domains are treated as regulatory, so that a linker that accepts a regulatory domain also accepts a binding or interaction domain;
\item $p_1\to p$ and $p_2\to p$: single- and multi-domain proteins are proteins;
\item $c\to p_1$ and $t\to p_1$: a catalytic or transporter domain can, on its own, constitute a complete single-domain protein. More generally, $x\to p_1$ is postulated for every domain type $x$ that occurs in single-domain proteins.
\end{enumerate}
Postulates (P2) and (P4) are modelling choices rather than biological facts. In particular, (P4) does not say that a domain \emph{is} a protein, only that a single word of type $x$ forms a complete ``sentence'', just as a one-word imperative does in English.

\paragraph{Linker types.} Linkers play the role of the relational words, such as transitive verbs, that connect the other words of a sentence. A linker between a domain of type $x$ and a domain of type $y$ is given the type
\begin{equation}\label{eq:linker1}
L_{xy} : \rt{x}\,p_2\,\lt{y},
\end{equation}
so that the two-domain protein $D_1L_1D_2$ with domain types $x$ and $y$ reduces as
\begin{equation}\label{eq:reduction2}
x\,(\rt{x}\,p_2\,\lt{y})\,y \;\to\; p_2 \;\to\; p,
\end{equation}
via the morphism $\epsilon^{r}_{x}\otimes\id_{p_2}\otimes\epsilon^{l}_{y}$, followed by $p_2\to p$. The linker lexicon contains an entry $L_{xy}$ for each ordered pair $(x,y)$ of domain types that is observed in real proteins. Unobserved pairs have no entry, so architectures that use them are ungrammatical. This makes the grammaticality check non-trivial: only a small fraction of all possible domain combinations occur in nature \citep{Apic2001,Yu2019}. When two domains are directly adjacent, the linker is a word with an empty sequence that still carries the type~\eqref{eq:linker1}. (In versions~1--3 the linker type was read off from its two neighbours for each protein, so that every two-domain protein was grammatical by construction.)

\paragraph{Proteins with more than two domains.} The type~\eqref{eq:linker1} alone does not suffice for $m\ge 3$ domains: in $x\,(\rt{x}p_2\lt{y})\,y\,(\rt{y}p_2\lt{z})\,z$ the middle domain is consumed by the first linker, and the string reduces only to $p_2\,\rt{y}\,p_2$, which is not a protein type. We therefore add continuation linkers
\begin{equation}\label{eq:linker2}
K_{z} : \rt{p_2}\,p_2\,\lt{z},
\end{equation}
which take the protein assembled so far on their left and the next domain on their right (Figure~\ref{fig:multidomain}).

\begin{proposition}\label{prop:multi}
Let $D_1L_1D_2L_2\cdots L_{m-1}D_m$ $(m\ge2)$ be a protein whose domains have types $x_1,\ldots,x_m$. Typing $L_1$ as $L_{x_1x_2}:\rt{x_1}p_2\lt{x_2}$ and $L_k$ $(2\le k\le m-1)$ as $K_{x_{k+1}}:\rt{p_2}p_2\lt{x_{k+1}}$, the type of the protein reduces to $p_2$, and hence to $p$.
\end{proposition}
\begin{proof}
By induction on $m$. For $m=2$ this is~\eqref{eq:reduction2}. If the prefix $D_1L_1\cdots D_{k}$ reduces to $p_2$, then the prefix $D_1\cdots D_{k+1}$ reduces as $p_2\,(\rt{p_2}\,p_2\,\lt{x_{k+1}})\,x_{k+1}\to p_2$ by $\epsilon^{r}_{p_2}\otimes\id_{p_2}\otimes\epsilon^{l}_{x_{k+1}}$.
\end{proof}

\begin{figure}[htbp]
\centering
\begin{tikzpicture}[x=1cm,y=1cm,every node/.style={font=\small}]
\node (a) at (0,0) {$x$};
\node (b) at (1.3,0) {$\rt{x}$};
\node (c) at (2.0,0) {$p_2$};
\node (d) at (2.7,0) {$\lt{y}$};
\node (e) at (4.0,0) {$y$};
\node (f) at (5.3,0) {$\rt{p_2}$};
\node (g) at (6.0,0) {$p_2$};
\node (h) at (6.7,0) {$\lt{z}$};
\node (i) at (8.0,0) {$z$};
\draw[thick] (a.south) to[out=-90,in=-90,looseness=1.2] (b.south);
\draw[thick] (d.south) to[out=-90,in=-90,looseness=1.2] (e.south);
\draw[thick] (c.south) to[out=-90,in=-90,looseness=1.0] (f.south);
\draw[thick] (h.south) to[out=-90,in=-90,looseness=1.2] (i.south);
\draw[thick] (g.south) -- ++(0,-1.6) node[below] {$p_2\to p$};
\draw[decorate,decoration={brace,amplitude=4pt}] (-0.25,0.35) -- (0.25,0.35) node[midway,above=4pt] {$D_1$};
\draw[decorate,decoration={brace,amplitude=4pt}] (1.0,0.35) -- (3.0,0.35) node[midway,above=4pt] {$L_1=L_{xy}$};
\draw[decorate,decoration={brace,amplitude=4pt}] (3.75,0.35) -- (4.25,0.35) node[midway,above=4pt] {$D_2$};
\draw[decorate,decoration={brace,amplitude=4pt}] (5.0,0.35) -- (7.0,0.35) node[midway,above=4pt] {$L_2=K_{z}$};
\draw[decorate,decoration={brace,amplitude=4pt}] (7.75,0.35) -- (8.25,0.35) node[midway,above=4pt] {$D_3$};
\end{tikzpicture}
\caption{Pregroup reduction of a three-domain protein $D_1L_1D_2L_2D_3$ with domain types $x,y,z$. Each cup is a contraction; the first linker assembles $D_1$ and $D_2$ into a multi-domain protein ($p_2$), which the continuation linker $L_2$ extends with $D_3$. The one remaining wire carries the protein type.}
\label{fig:multidomain}
\end{figure}

One can check that the reduction in Proposition~\ref{prop:multi} is the only reduction of these types to a protein type. The left-to-right (incremental) bracketing is a modelling choice: other bracketings, for example ones that reflect contacts between non-adjacent domains in the folded protein, could be encoded by other lexical types and are left to future work.

\begin{remark}[Terminal regions]\label{rem:terminal}
Versions~1--3 ignored the N- and C-terminal regions. This can be made precise by typing them as modifiers of the protein, $\mathrm{NTR}: p\,\lt{p}$ and $\mathrm{CTR}: \rt{p}\,p$, in the same way as adjectives modify nouns. Then $(p\lt{p})\,p_2\,(\rt{p}p)\to p$, and the meaning of each terminal region is a linear map $P\to P$. Ignoring a terminal region amounts to setting this map to the identity. Terminal regions can carry functional signals (for example targeting signals or disordered regulatory tails), so non-identity maps may be needed in practice.
\end{remark}

\subsection{Meaning spaces and vector representations of domains}\label{sec:vectors}

\paragraph{Meaning spaces.} Following Section~\ref{sec:lifting}, we fix a strong monoidal functor $F$ that sends
\begin{itemize}[nosep]
\item every domain type ($b$, $b_k$, $c$, $c_j$, $i$, $r$, $t$) to one domain space $D$;
\item every protein type ($p$, $p_1$, $p_2$) to one protein space $P$, so that all proteins have meanings in the same space and can be compared by~\eqref{eq:similarity}, just as all sentences are in the DisCoCat model;
\item every order relation among domain types, and among protein types, to the identity map, and each relation $x\to p_1$ in (P4) to a linear map $\pi: D\to P$.
\end{itemize}
Using a single space $D$ for all domain types is natural when all domain vectors come from one embedding method, as below. The types still matter, because they decide which linker tensors may combine which domains. Two choices of $P$ are natural: $P=D$ with $\pi=\id_D$, so that composed protein vectors are directly comparable with embeddings of whole protein sequences; or a space of functional annotations, for example with one basis vector per Gene Ontology (GO) term \citep{Ashburner2000}, which makes protein meanings interpretable.

Each word then has a meaning space determined by its type: a domain has a vector $u\in D$; a first linker $L_{xy}:\rt{x}p_2\lt{y}$ has a tensor in $D\otimes P\otimes D$; a continuation linker $K_z:\rt{p_2}p_2\lt{z}$ has a tensor in $P\otimes P\otimes D$.

\paragraph{Distributional representations of domains.} The idea behind distributional representations of words is captured by Firth's dictum ``you shall know a word by the company it keeps'': words that occur in similar contexts tend to have similar meanings \citep{Firth1957,Coecke2010}. In count-based models, a word is represented by a vector whose basis vectors correspond to context words \citep{Coecke2010}; in prediction-based models such as Word2Vec, words are represented by dense vectors learned by a neural network \citep{Hinton1986,Mikolov2013}. Both kinds of vector can serve as word meanings in the DisCoCat model.

\begin{figure}[htbp]
\centering
\includegraphics[width=0.45\textwidth]{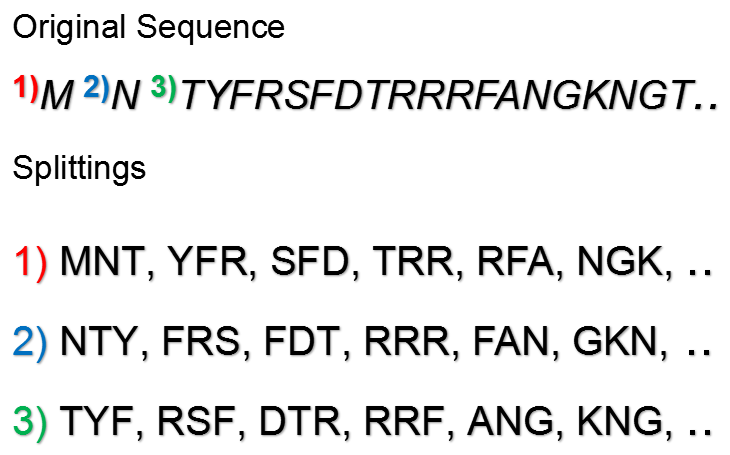}
\caption{Splitting of a protein sequence into three lists of non-overlapping 3-mers in ProtVec \citep{Asgari2015}.}
\label{fig:protvec}
\end{figure}

Distributional representations of proteins are equally desirable. Because vectors are easy for computers to handle, they can serve as general-purpose features for many bioinformatics problems, such as protein family classification, structure prediction and visualization. ProtVec \citep{Asgari2015} applies Word2Vec to protein sequences. It was trained on 546{,}790 manually reviewed Swiss-Prot sequences. Each sequence is split into three lists of non-overlapping 3-mers, one for each of the three possible starting positions (Figure~\ref{fig:protvec}). Each 3-mer is treated as a ``word'' and is assigned a 100-dimensional vector by Word2Vec \citep{Mikolov2013}. A sequence is then represented by the sum of the vectors of its overlapping 3-mers \citep{Asgari2015}. With these representations, ProtVec classified 324{,}018 Swiss-Prot sequences from 7{,}027 protein families with an average accuracy of 93\%, outperforming existing methods, and it proved useful for visualizing protein data.

We obtain the vector $u\in D$ of a domain ($D=\mathbb{R}^{100}$) by applying ProtVec to the domain's subsequence, that is, by summing the vectors of the 3-mers that lie within the domain boundaries given by a domain database such as Pfam or InterPro \citep{ElGebali2019,Mitchell2019}. Domains are good ``words'' for this purpose because their sequences are conserved, whereas the ways in which domains are combined are highly variable \citep{Gimona2006}.

This also shows what the DisCoCat model adds. Summing the 3-mer vectors of a whole protein, as ProtVec does, gives (up to the 3-mers that span domain boundaries) the sum of its domain, linker and terminal-region vectors. Composition by summation is commutative and ignores the arrangement of domains: it is the ``bag-of-words'' baseline familiar from natural language processing \citep{Mitchell2010}. The DisCoCat model replaces it by a composition that depends on the types and order of the domains. Linkers are therefore \emph{not} represented by ProtVec vectors of their own (short, poorly conserved) sequences. Like transitive verbs, they are represented by tensors that encode how the two flanking domain functions are combined, and these tensors are learned from data (Section~\ref{sec:learning}).

\paragraph{Working hypotheses.} The construction rests on two hypotheses, which should be tested rather than assumed:
\begin{enumerate}[label=(H\arabic*),nosep]
\item domains that occur in similar sequence contexts have similar functions, so domain vectors reflect domain function (the distributional hypothesis for domains);
\item the function of a protein is, to a useful approximation, a multilinear function of the functions of its domains, determined by its domain architecture (the compositional hypothesis).
\end{enumerate}

\subsection{Composing the function of a protein}\label{sec:composing}

We first illustrate the construction with a toy example of a hypothetical protein, and then apply it to real proteins.

\begin{figure}[htbp]
\centering
\includegraphics[width=0.6\textwidth]{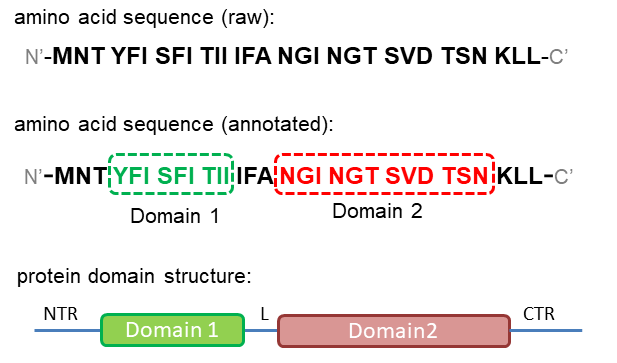}
\caption{Sequence and domain structure of the toy protein mockP. NTR, N-terminal region; L, linker; CTR, C-terminal region. In this and the following figures, colours distinguish the domains within a protein and do not indicate their types.}
\label{fig:mockP}
\end{figure}

\paragraph{A toy example.} Consider a hypothetical protein, mockP, with the amino acid sequence and domain structure shown in Figure~\ref{fig:mockP}. Domain1 is a regulatory domain and Domain2 a catalytic domain. (Real proteins have much longer sequences.)

mockP has five components: NTR, Domain1, linker L, Domain2 and CTR. Ignoring the terminal regions (Remark~\ref{rem:terminal}), the types are
\[
\text{Domain1}: r,\qquad L = L_{rc}: \rt{r}\,p_2\,\lt{c},\qquad \text{Domain2}: c,
\]
and the type of mockP reduces as $r\,(\rt{r}p_2\lt{c})\,c\to p_2$ via $\epsilon^{r}_{r}\otimes\id_{p_2}\otimes\epsilon^{l}_{c}$. Applying $F$, the meaning spaces are $D\otimes(D\otimes P\otimes D)\otimes D$, and the reduction becomes the linear map
\[
f = \epsilon_D\otimes\id_P\otimes\epsilon_D : D\otimes(D\otimes P\otimes D)\otimes D\to P .
\]
Writing $\{e_i\}$ for an orthonormal basis of $D$, $\{f_j\}$ for one of $P$, and $L=\sum_{ijk}L_{ijk}\,e_i\otimes f_j\otimes e_k$, the meaning of a two-domain protein with domain vectors $u$ (left) and $w$ (right) is
\begin{equation}\label{eq:twodomain}
\vec{v} = f(u\otimes L\otimes w) = \sum_{j}\Bigl(\sum_{i,k}u_i\,L_{ijk}\,w_k\Bigr) f_j = \sum_j \bigl(u^{\top}L^{(j)}w\bigr) f_j ,
\end{equation}
where $L^{(j)}$ is the matrix with entries $L_{ijk}$ for fixed $j$. Each coordinate of the protein vector is thus a bilinear form in the two domain vectors (Figure~\ref{fig:stringdiagram}).

\begin{figure}[htbp]
\centering
\begin{tikzpicture}[x=1cm,y=1cm,every node/.style={font=\small}]
\draw[thick,fill=green!15] (-0.6,0) -- (0.6,0) -- (0,-0.8) -- cycle; \node at (0,-0.3) {$u$};
\draw[thick,fill=gray!15] (1.6,0) -- (4.4,0) -- (3,-0.8) -- cycle; \node at (3,-0.3) {$L_{rc}$};
\draw[thick,fill=red!15] (5.4,0) -- (6.6,0) -- (6,-0.8) -- cycle; \node at (6,-0.3) {$w$};
\draw[thick] (0,-0.8) -- (0,-1.5) to[out=-90,in=-90,looseness=1.3] (2.2,-1.5) -- (2.2,-0.5);
\draw[thick] (3.8,-0.5) -- (3.8,-1.5) to[out=-90,in=-90,looseness=1.3] (6,-1.5) -- (6,-0.8);
\draw[thick] (3,-0.8) -- (3,-3.0) node[below] {$P$};
\node[left] at (0,-1.2) {$D$};
\node[right] at (2.2,-1.1) {$D$};
\node[left] at (3.8,-1.1) {$D$};
\node[right] at (6,-1.2) {$D$};
\node at (1.1,-2.55) {$\epsilon_D$};
\node at (4.9,-2.55) {$\epsilon_D$};
\node[anchor=west] at (7.2,-1.5) {$=\ (\epsilon_D\otimes\id_P\otimes\epsilon_D)(u\otimes L_{rc}\otimes w)\in P$};
\end{tikzpicture}
\caption{String diagram of the composition~\eqref{eq:twodomain} for mockP. Triangles are the meaning vectors of the words (Domain1 $u$, linker $L_{rc}$, Domain2 $w$); cups are the maps $\epsilon_D$ (inner products) that lift the pregroup contractions; the single open wire carries the protein vector in $P$.}
\label{fig:stringdiagram}
\end{figure}

For a numerical illustration, take $D=\mathbb{R}^{3}$ (instead of the 100 dimensions of ProtVec) and $P=\mathbb{R}^{2}$, with
\[
u=\text{Domain1}=\begin{pmatrix}30\\12\\9\end{pmatrix},\quad
w=\text{Domain2}=\begin{pmatrix}2\\4\\10\end{pmatrix},\quad
L^{(1)}=\begin{pmatrix}1&0&0\\0&1&0\\0&0&1\end{pmatrix},\quad
L^{(2)}=\begin{pmatrix}0&1&0\\0&0&1\\1&0&0\end{pmatrix}.
\]
Then~\eqref{eq:twodomain} gives
\begin{equation}\label{eq:mockP}
\vec v_{\text{mockP}} = \bigl(u^{\top}L^{(1)}w\bigr)f_1+\bigl(u^{\top}L^{(2)}w\bigr)f_2 = 198\,f_1 + 258\,f_2 = \begin{pmatrix}198\\258\end{pmatrix}\in P .
\end{equation}

Two points are worth noting. First, in versions~1--3 the linker was taken to be the $3\times 3$ identity matrix, an order-two tensor, so the protein wire $P$ was missing and the ``protein meaning'' was the scalar $\ip{u}{w}=198$. That number is the inner product of the two domain vectors, a similarity between the domains, not a representation of the protein's function. It is the special case $P=\mathbb{R}$, $L=\sum_i e_i\otimes 1\otimes e_i$ of~\eqref{eq:twodomain}, in which all proteins would have one-dimensional meanings. Second, the composition depends on the arrangement of the domains. Applying the same tensor with the two domains swapped gives $(198,360)^{\top}$ rather than $(198,258)^{\top}$, because $L^{(2)}$ is not symmetric. The additive baseline $u+w=(32,16,19)^{\top}$ cannot distinguish the two orders.

\paragraph{A three-domain toy example.} Suppose a third domain $z=(1,0,2)^{\top}$ of type $b$ follows mockP after a continuation linker $K_{b}:\rt{p_2}p_2\lt{b}$ with tensor $K=\sum_{ajk}K_{ajk}f_a\otimes f_j\otimes e_k\in P\otimes P\otimes D$. By Proposition~\ref{prop:multi}, the meaning of the three-domain protein is
\[
\vec v' = (\epsilon_P\otimes\id_P\otimes\epsilon_D)(\vec v_{\text{mockP}}\otimes K\otimes z) = \sum_j\Bigl(\sum_{a,k}(\vec v_{\text{mockP}})_a K_{ajk} z_k\Bigr)f_j .
\]
With $K^{(1)}=\left(\begin{smallmatrix}1&0&0\\0&0&1\end{smallmatrix}\right)$ and $K^{(2)}=\left(\begin{smallmatrix}0&1&0\\1&0&0\end{smallmatrix}\right)$ (rows indexed by $a$, columns by $k$), we obtain $\vec v'=(714,258)^{\top}$, again a vector in $P$. Its cosine similarity~\eqref{eq:similarity} with $\vec v_{\text{mockP}}$ is $0.84$.

\paragraph{Real proteins.} We now turn to real proteins from the fruit fly \textit{Drosophila melanogaster}. We chose fruit-fly rather than human proteins for three reasons: (1) the fruit fly is the model organism used in the author's research; (2) fruit-fly genetics has been studied extensively for over a century, so its genome is among the best characterized of all multicellular eukaryotes \citep{ModENCODE2019,Hales2015}; and (3) at the molecular level, and particularly for functional domains, fruit-fly and human proteins are highly conserved \citep{Hales2015,Pandey2011}. Domain names and annotations are from InterPro \citep{Mitchell2019}. The types, reductions and composed meanings are summarized in Table~\ref{tab:proteins}.

\begin{table}[htbp]
\centering
\small
\caption{Types, reductions and composed meanings of five \textit{Drosophila} proteins. $u_\bullet$ denotes the ProtVec vector in $D$ of the named domain; $L^{(j)}_{xy}$ are the slices of the linker tensor for the architecture $(x,y)$; $\{f_j\}$ is a basis of $P$.}
\label{tab:proteins}
\begin{tabular}{@{}l l l l@{}}
\toprule
Protein & Domain types & Reduction & Meaning in $P$\\
\midrule
DAT  & $t$ & $t\to p_1\to p$ & $\pi(u_{\mathrm{SNF}})$\\[0.5ex]
FoxP & $r$, $b_2$ & $r\,(\rt{r}p_2\lt{b_2})\,b_2\to p_2$ & $\sum_j (u_{\mathrm{CC}}^{\top}L^{(j)}_{rb_2}u_{\mathrm{FH}})\,f_j$\\[0.5ex]
p53  & $b_2$, $r$ & $b_2\,(\rt{b_2}p_2\lt{r})\,r\to p_2$ & $\sum_j (u_{\mathrm{DBD}}^{\top}L^{(j)}_{b_2r}u_{\mathrm{CTD}})\,f_j$\\[0.5ex]
Fru  & $r$, $b_2$ & $r\,(\rt{r}p_2\lt{b_2})\,b_2\to p_2$ & $\sum_j (u_{\mathrm{BTB}}^{\top}L^{(j)}_{rb_2}u_{\mathrm{ZF}})\,f_j$\\[0.5ex]
Sxc  & $i$, $c_2$ & $i\,(\rt{i}p_2\lt{c_2})\,c_2\to p_2$ & $\sum_j (u_{\mathrm{TPR}}^{\top}L^{(j)}_{ic_2}u_{\mathrm{OGT}})\,f_j$\\
\bottomrule
\end{tabular}
\end{table}

\begin{figure}[htbp]
\centering
\includegraphics[width=0.5\textwidth]{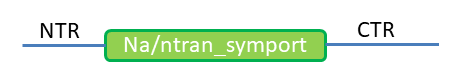}
\caption{Domain structure of DAT (dopamine transporter): a single sodium:neurotransmitter symporter domain (Na/ntran\_symport).}
\label{fig:DAT}
\end{figure}

\emph{DAT.} Many proteins consist of a single domain, while others have several. We begin with the single-domain protein DAT (dopamine transporter) of the fly. Its transporter function is carried out by a sodium:neurotransmitter symporter domain (SNF; ``Na/ntran\_symport'' in Figure~\ref{fig:DAT}), which we type as $t$. By postulates (P3) and (P4), $t\to p_1\to p$, so DAT is a grammatical protein. No linker is involved, and its meaning is $\pi(u_{\mathrm{SNF}})\in P$, where $u_{\mathrm{SNF}}$ is the ProtVec vector of the domain; for $P=D$, $\pi$ is the identity and this is just $u_{\mathrm{SNF}}$.

\begin{figure}[htbp]
\centering
\includegraphics[width=0.5\textwidth]{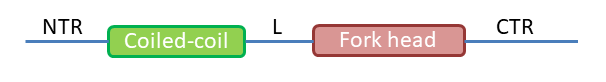}
\caption{Domain structure of FoxP: a coiled-coil region and a fork-head DNA-binding domain separated by a linker (L).}
\label{fig:FoxP}
\end{figure}

\emph{FoxP.} The second example is FoxP, a transcription factor of the Forkhead box P (FoxP) subfamily, with the domain structure shown in Figure~\ref{fig:FoxP}. Its human orthologues are FOXP1--FOXP4; mutations in FOXP2 cause a severe speech and language disorder \citep{Lai2001}. According to InterPro \citep{Mitchell2019}, the coiled-coil (CC) region modulates the dimerization of FoxP transcription factors, and the fork-head (FH) domain is a conserved DNA-binding domain. We type them as $r$ and $b_2$, so the linker is $L_{rb_2}:\rt{r}p_2\lt{b_2}$. The type $r(\rt{r}p_2\lt{b_2})b_2$ reduces to $p_2$ via $\epsilon^{r}_{r}\otimes\id_{p_2}\otimes\epsilon^{l}_{b_2}$, which is lifted to $\epsilon_D\otimes\id_P\otimes\epsilon_D$ on $D\otimes(D\otimes P\otimes D)\otimes D$. By~\eqref{eq:twodomain}, the meaning of FoxP is $\sum_j(u_{\mathrm{CC}}^{\top}L^{(j)}_{rb_2}u_{\mathrm{FH}})f_j$.

\begin{figure}[htbp]
\centering
\includegraphics[width=0.5\textwidth]{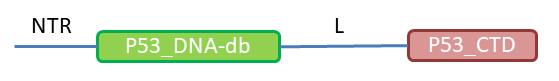}
\caption{Domain structure of \textit{Drosophila} p53: a DNA-binding domain (P53\_DNA-db) and a C-terminal regulatory domain (P53\_CTD).}
\label{fig:p53}
\end{figure}

\emph{p53.} The third example is p53. In humans p53 is a well-known tumour suppressor, and in \textit{Drosophila} it plays important roles in regulating cell death \citep{Levine1997,Jin2000,Sogame2003}. Fly p53 is a transcription factor with a DNA-binding domain (DBD; ``P53\_DNA-db'') and a C-terminal regulatory domain (CTD; ``P53\_CTD'') (Figure~\ref{fig:p53}). Here the DNA-binding domain comes first, so the linker is $L_{b_2r}:\rt{b_2}p_2\lt{r}$, a different lexical entry from that of FoxP. The type $b_2(\rt{b_2}p_2\lt{r})r$ reduces to $p_2$ via $\epsilon^{r}_{b_2}\otimes\id_{p_2}\otimes\epsilon^{l}_{r}$, and the meaning of p53 is $\sum_j(u_{\mathrm{DBD}}^{\top}L^{(j)}_{b_2r}u_{\mathrm{CTD}})f_j$.

\begin{figure}[htbp]
\centering
\includegraphics[width=0.5\textwidth]{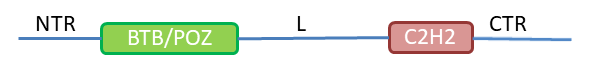}
\caption{Domain structure of Fruitless (Fru): a BTB/POZ domain and a C2H2 zinc-finger domain.}
\label{fig:Fru}
\end{figure}

\emph{Fruitless.} Fruitless (Fru) is involved in sex determination and the control of male courtship behaviour \citep{Ryner1996}. It has a BTB/POZ (Broad-complex, Tramtrack and Bric-\`a-brac / Poxvirus and zinc finger) domain, which we type as regulatory ($r$), and a C2H2 (Cys$_2$His$_2$) zinc-finger (ZF) DNA-binding domain ($b_2$) (Figure~\ref{fig:Fru}). Fru therefore has the same architecture $(r,b_2)$ as FoxP and uses the same linker entry $L_{rb_2}$; its meaning is $\sum_j(u_{\mathrm{BTB}}^{\top}L^{(j)}_{rb_2}u_{\mathrm{ZF}})f_j$. In the model, FoxP and Fru differ only through their domain vectors, and their functional similarity is given by~\eqref{eq:similarity} applied to the two composed vectors.

\begin{figure}[htbp]
\centering
\includegraphics[width=0.5\textwidth]{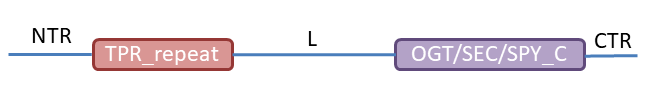}
\caption{Domain structure of Super sex combs (Sxc): a tetratricopeptide repeat region (TPR\_repeat) and an OGT/SEC/SPY C-terminal catalytic domain.}
\label{fig:Sxc}
\end{figure}

\emph{Super sex combs.} Finally, Super sex combs (Sxc) is the fly O-linked $N$-acetylglucosamine transferase (OGT), a glycosyltransferase that is essential for Polycomb repression and hence for the control of development \citep{Gambetta2009}. Sxc contains a tetratricopeptide repeat (TPR) region, which mediates interactions with other proteins ($i$), and the C-terminal catalytic domain of the OGT/SEC/SPY (OGT, SECRET AGENT and SPINDLY) family (``OGT/SEC/SPY\_C''), which carries the transferase activity ($c_2$) (Figure~\ref{fig:Sxc}). The linker is $L_{ic_2}:\rt{i}p_2\lt{c_2}$, the type $i(\rt{i}p_2\lt{c_2})c_2$ reduces to $p_2$ via $\epsilon^{r}_{i}\otimes\id_{p_2}\otimes\epsilon^{l}_{c_2}$, and the meaning of Sxc is $\sum_j(u_{\mathrm{TPR}}^{\top}L^{(j)}_{ic_2}u_{\mathrm{OGT}})f_j$.

\subsection{Comparing proteins and learning linker tensors}\label{sec:learning}

All composed protein meanings lie in the same space $P$, whatever the number and types of their domains. The functional similarity of two proteins can therefore be measured by~\eqref{eq:similarity}, for example between FoxP and Fru, or between a single-domain and a multi-domain protein.

The domain vectors come from ProtVec, but the linker tensors must be estimated. Because~\eqref{eq:twodomain} is linear in $L$, each slice $L^{(j)}_{xy}$ can be estimated by (regularized) least-squares regression from proteins of architecture $(x,y)$ whose function vectors $t$ are known, for example from GO annotations: minimize $\sum_k\lVert f(u_k\otimes L_{xy}\otimes w_k)-t_k\rVert^2$ over the proteins $k$ of that architecture. This follows the regression methods for relational words in natural language \citep{Baroni2010,Grefenstette2013}. A simpler, unsupervised starting point, in the spirit of \citet{Grefenstette2011}, is the relational tensor $L_{xy}=\sum_k u_k\otimes t_k\otimes w_k$ summed over the same proteins. Continuation linkers $K_z$ are estimated in the same way from proteins with three or more domains. Because there are far fewer architectures than proteins, each linker tensor is shared by many proteins, and it applies to new proteins with the same architecture but different domains.

\section{Related developments since 2019}\label{sec:recent}

Several developments since the first version of this paper bear on the model.

\paragraph{Protein language models.} Large transformer-based protein language models trained on millions of sequences, such as ESM-1b \citep{Rives2021}, the ProtTrans models \citep{Elnaggar2022} and ESM-2 \citep{Lin2023}, learn contextual embeddings of residues and proteins. Their representations encode structural and functional information, and ESM-2 even supports atomic-level structure prediction from a single sequence \citep{Lin2023}. These models can supply much richer domain vectors than ProtVec, for example by averaging residue embeddings over a domain's boundaries. They also weaken the argument, made in versions~1--3, that only domains, and not whole proteins, can be given vector representations: whole-protein embeddings are now routine. The DisCoCat model remains complementary to them. Its composition is explicit and typed: the contribution of each domain and the role of the architecture can be read off from the linker tensors, and the model can be applied to new combinations of known domains. The embeddings of whole proteins from language models also provide natural targets and baselines for testing it (Section~\ref{sec:discussion}). For a review of natural language processing methods applied to proteins, see \citet{Ofer2021}.

\paragraph{Domains as words.} Two studies have treated domains explicitly as words. \citet{Yu2019} applied $n$-gram analysis to the domain architectures of proteins across the tree of life. They found that the information gain of real architectures over random domain combinations is nearly constant, at about 1.2~bits, which suggests a ``quasi-universal grammar'' of protein domains. This supports the use of a linker lexicon that admits only attested domain combinations (Section~\ref{sec:typing}). The dom2vec method \citep{Melidis2021} learns embeddings of InterPro domains from domain architectures, treating proteins as sentences of domains. These embeddings are an alternative source of domain vectors in $D$ that, unlike ProtVec, are based on the contexts of domains rather than on their sequences.

\paragraph{Developments of the DisCoCat model.} The DisCoCat model has since been given functorial formulations for the Lambek calculus as well as pregroups \citep{Coecke2013}, has been implemented in software, notably the lambeq library, which turns sentences into string diagrams and tensor networks \citep{Kartsaklis2021}, and has been run on quantum hardware, where syntax-sensitive models were compared with bag-of-words baselines \citep{Lorenz2023}. DisCoCirc extends the framework from sentences to texts, in which the meanings of entities evolve as sentences act on them \citep{Coecke2021}. A protein analogue of this idea might describe how the proteins of a pathway or complex modify one another's functions. Tools such as lambeq could be used to implement the protein model of Section~\ref{sec:protein}.

\paragraph{Function annotation.} The Gene Ontology \citep{Ashburner2000} and the InterPro2GO mapping of domains to GO terms \citep{Burge2012} supply function annotations at the level of both proteins and domains. They can be used to construct the protein space $P$, to train the linker tensors, and to evaluate the model with established measures of semantic similarity \citep{Pesquita2009}.

\section{Discussion: limitations and how to test the model}\label{sec:discussion}

The model rests on simplifying assumptions. Domain boundaries are taken from annotation databases; the order of the domains is the only syntax considered, so interactions between domains that are not adjacent in sequence but in contact in the folded structure are not modelled; post-translational modifications, alternative splicing and the formation of complexes are ignored; and the composition is multilinear. The broad functional classes used as types are a first proposal and can be refined or replaced, for example by types derived from domain families. Hypotheses (H1) and (H2) are empirical claims that must be tested.

A test could proceed as follows. (i) Collect proteins with annotated domain architectures, for example from InterPro, together with their GO annotations. (ii) Compute domain vectors with ProtVec, dom2vec or a protein language model, and construct protein target vectors from GO annotations. (iii) Estimate the linker tensors by regression (Section~\ref{sec:learning}) on a training set. (iv) On held-out proteins, compare the cosine similarities~\eqref{eq:similarity} of composed vectors with GO-based functional similarities \citep{Pesquita2009}, and compare the model with simpler baselines: the sum and the element-wise product of the domain vectors \citep{Mitchell2010}, and embeddings of whole proteins. Of particular interest is how well the model generalizes to architectures and domain combinations not seen during training.

\section{Conclusion and future work}

In this work we applied DisCoCat, a mathematical model built for natural language, to protein linguistics. We gave pregroup types to domains and linkers; lifted the type reductions to multilinear maps; showed that every protein, whatever its number of domains, then receives a meaning vector in a single protein space; and demonstrated the procedure with numerical toy examples and five \textit{Drosophila} proteins. The result is a formal, compositional representation of protein function in a vector space.

Our main aim has been to show that a state-of-the-art theory from natural language processing can be applied to the language of nature. To make the method useful in practice, three issues must be addressed. First, the grammatical types proposed here are preliminary and need to be refined and validated against data. Second, the linker tensors must be learned from data; this, rather than the domain vectors, is the central practical question. Third, the model must be tested against simpler compositional baselines and against protein language models, as outlined in Section~\ref{sec:discussion}.

The diagrammatic calculus of the DisCoCat model \citep{Coecke2010}, illustrated in Figures~\ref{fig:multidomain} and~\ref{fig:stringdiagram}, makes the composition of protein functions intuitive and simplifies calculations. We expect it to be equally useful for the extensions discussed above, such as non-sequential domain contacts and interactions between proteins.

\section*{Acknowledgements}
I thank Quanlong Wang (Department of Computer Science, University of Oxford), co-author of versions~1--3 of this paper, for his contributions to the development of this work. The revision for version~4 was prepared with the assistance of Claude Opus~5.5 (Anthropic), an AI model, which was used to check the earlier versions against the cited literature, identify errors and logical gaps, draft corrections and new text, and typeset the manuscript. The author reviewed all changes and takes full responsibility for the content.

\bibliographystyle{apalike}
\bibliography{references_V4}

\end{document}